\makeatletter
\declare@file@substitution{revtex4-1.cls}{revtex4-2.cls}
\makeatother

\documentclass{aastex631}

\begin{document}

\title{Solar Magnetic Polarity Effect on Neutron Monitor Count Rates: Comparing Latitude Surveys and Antarctic Stations}

\author[0000-0002-8959-3273]{K. Poopakun}
\affiliation{ Graduate Program in Astronomy, Department of Physics and Materials Science, \\Faculty of Science, Chiang Mai University, Chiang Mai 50200, Thailand}

\author[0000-0002-1664-5845]{W. Nuntiyakul}
\affiliation{Department of Physics and Materials Science, \\Faculty of Science, Chiang Mai University, Chiang Mai 50200, Thailand}

\author[0009-0008-8182-1180]{S. Khamphakdee}
\affiliation{Department of Physics and Materials Science, \\Faculty of Science, Chiang Mai University, Chiang Mai 50200, Thailand}

\author[0000-0002-3764-8949]{A. Seripienlert}
\affiliation{National Astronomical Research Institute of Thailand (NARIT), Chiang Mai 50180, Thailand}


\author[0000-0003-3414-9666]{D. Ruffolo}
\affiliation{Department of Physics, Faculty of Science, Mahidol University, Bangkok 10400, Thailand}

\author[0000-0001-7929-810X]{P. Evenson}
\affiliation{Department of Physics and Astronomy, University of Delaware, Newark, DE 19716, USA}

\author{P. Jiang}
\affiliation{Polar Research Institute of China, Pudong, Shanghai 200136, China}

\author[0000-0001-6327-1113]{P. Chuanraksasat}
\affiliation{National Astronomical Research Institute of Thailand (NARIT), Chiang Mai 50180, Thailand}

\author[0000-0002-2131-4100]{K. Munakata}
\affiliation{Physics Department, Shinshu University, Matsumoto, Nagano 390-8621, Japan}

\author[0000-0001-7463-8267]{M. L. Duldig}
\affiliation{School of Natural Sciences, University of Tasmania, Hobart, Tasmania 7001, Australia}

\author[0000-0002-4698-1672]{J. E. Humble}
\affiliation{School of Natural Sciences, University of Tasmania, Hobart, Tasmania 7001, Australia}

\author[0000-0003-2415-9959]{J. Madsen}
\affiliation{Wisconsin IceCube Particle Astrophysics Center, University of Wisconsin-Madison, WI 53703, USA}

\author[0000-0001-5511-7183]{B. Soonthornthum}
\affiliation{National Astronomical Research Institute of Thailand (NARIT), Chiang Mai 50180, Thailand}

\author[0000-0001-7987-017X]{S. Komonjinda}
\affiliation{Department of Physics and Materials Science, \\Faculty of Science, Chiang Mai University, Chiang Mai 50200, Thailand}


\begin{abstract}
The Galactic cosmic ray spectrum manifests pronounced variations over the 11-year sunspot cycle and more subtle variations over the 22-year solar magnetic cycle. An important tool to study these variations is repeated latitude surveys with neutron monitors (NMs) onboard icebreakers in conjunction with land-based references. We revisit 13 annual latitude surveys from 1994 to 2007 using reference data from the Mawson NM instead of McMurdo NM (which closed in 2017). We then consider two more latitude surveys (2018 and 2019) with a monitor similar to the 3NM64 in the previous surveys but without lead rings around the central tube, a so-called ``semi-leaded neutron monitor.''  The new surveys extend the linear relationship among data taken at different cutoff rigidity ranges. They also confirm the ``crossover'' measured near solar minima during epochs of opposite solar magnetic polarity and the absence of a crossover for epochs having the same solar magnetic polarity.

\end{abstract}



\section{Introduction} \label{sec:intro}
Galactic cosmic rays (GCRs) are high-energy particles originating from outside the solar system that arrive at Earth. As GCRs are an important component of space weather, their impacts on Earth's atmosphere and technology need to be understood. GCRs can cause damage to electronics and communication systems and pose a health risk to astronauts and airline crews~\citep{Wilson2003}. As GCRs enter the heliosphere, they encounter a turbulent magnetic field, leading to significant variations in their intensity and energy levels~\citep{Parker1965}. The solar influence on GCR intensity is known as solar modulation. The GCR spectrum undergoes changes that are closely tied to both the sunspot cycle and the solar magnetic cycle~\citep{Forbush1954,JokipiiandThomas1981}.

\begin{figure}
	\centering
	\includegraphics[width=1.0\textwidth]{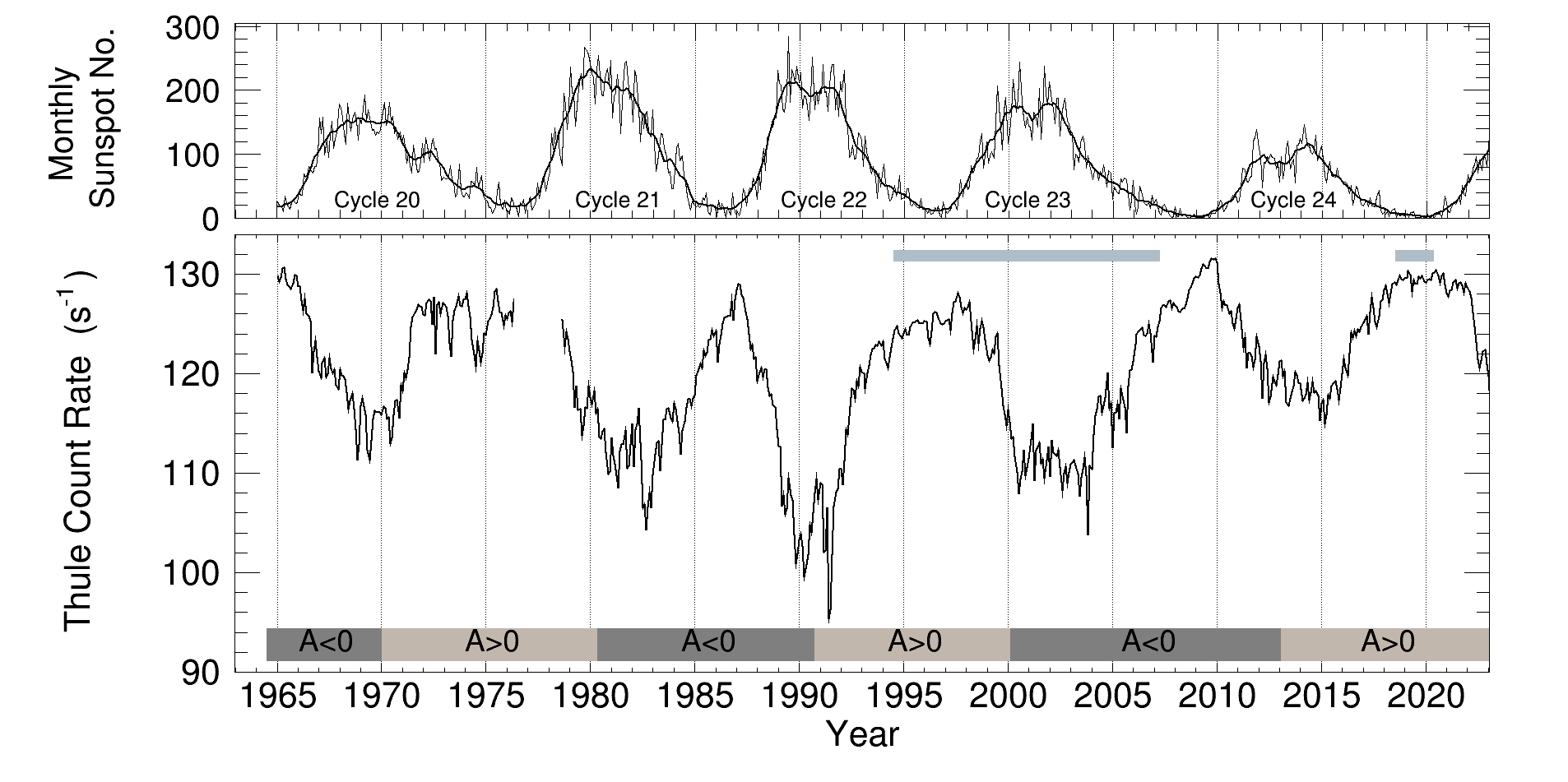}
	\caption{The count rate recorded by a polar neutron monitor (NM) indicates the Galactic cosmic ray flux above $\sim$1 GeV nucleon$^{-1}$ as it undergoes ``solar modulation'' related to solar activity. As the raw or 13-month smoothed monthly sunspot number, which is an indicator of solar activity, increases (shown in the top panel,  Source: WDC-SILSO Royal Observatory of Belgium, Brussels), the pressure-corrected monthly count rate recorded by the  Thule NM decreases (bottom panel, Source: Bartol Research Institute, University of Delaware, USA). The solar magnetic polarity reversal can be observed as the polarity shifts between positive (represented by $\mathrm{A}>0$) and negative (represented by $\mathrm{A}<0$) values. This work presents latitude survey observations for the periods 1994-2007 and 2019-2020, as indicated by horizontal bars between the two panels.}
	\label{fig:solar_modulation}
\end{figure}

An extensive record of cosmic ray intensity is derived from ground-level observations using neutron monitors (NMs) and muon detectors. Figure~\ref{fig:solar_modulation} provides a clear illustration of this phenomenon. The top panel displays the monthly sunspot number, an indicator of solar activity, as well as a smoothed monthly sunspot number. The smoothed value is derived from the 13-month smoothed sunspot number calculated using a ``tapered-boxcar'' moving average over a 13-month period, centered on the data for the respective month. The bottom panel shows an NM count rate, which is evidently related to solar activity, exhibiting an 11-year cycle. This quasi-periodicity is reflected in sunspot records dating back to the early 1600s and in the intensity of GCRs observed more directly since the 1950s. Variability on even longer timescales is also evident in studies of tree rings, ice cores, and sea sediments~\citep{RigozoEA08,YamaguchiEA10}.

The 11-year solar activity cycle, which is most clearly evident in the number of sunspots, is driven by the evolution of the solar magnetic field. With increased solar activity, the magnetic fluctuations in the solar wind become stronger and more effective in removing GCR from the inner solar system. This, in turn, leads to a decrease in count rates since NMs respond to Galactic cosmic rays in the GeV energy range \citep{Carimichael1964}. The polarity of the large-scale solar magnetic field reverses approximately every 11 years~\citep{ThambyahpillaiElliot1953}. The exact cause of these cycles and reversals remains unknown, but the large-scale reversal occurs when the magnetic field is most complex, near solar maximum.

A mobile neutron monitor rapidly collects data across various geomagnetic cutoffs during a ``latitude survey'' \citep{Rose1956,Moraal1989,NuntiyakulEA14,NuntiyakulEA18,NuntiyakulEA20}. These surveys use the geomagnetic cutoff rigidity, the minimum momentum per charge that particles must have to enter the atmosphere at a specific geographic location, to measure the differential response function, which relates to the primary cosmic ray spectrum. At a particular latitude and longitude, the cutoff rigidity for transmission through Earth's magnetic field is determined by the arrival direction of each primary cosmic ray, described by the local zenith and azimuthal angles~\citep{ClemEA97}.

Latitude surveys have been compared for times of similar solar modulation levels but opposite solar magnetic polarity. The surveys conducted in 1976 and 1987 \citep{Moraal1989} and 1997 and 2006 \citep{NuntiyakulEA14} demonstrate an intersection of differential response functions at a rigidity of approximately 5-7 GV. This intersection is called the ``crossover.''  \citet{Moraal1989} attributed this to gradient and curvature drift in the solar wind. When the magnetic field reverses, the drift patterns reverse, resulting in a different level of confinement in the heliosphere, leading to changes in the energy spectrum.  Alternatively \citet{BurgerEA97} 
noted that magnetic helicity could produce charge sign and magnetic polarity dependence of the diffusion coefficient, affecting solar modulation in the opposite direction, and suggested that the interaction of the polarity dependences of drifts and helicity could result in a crossover.

Crossovers are observed most clearly during relatively stable modulation periods at solar minimum. The transition from one state to the other was documented by \citet{NuntiyakulEA14} through the correlation of NM count rates from latitude surveys and a fixed NM located at McMurdo Station.  Regression analysis revealed a consistent linear trend with slopes that change rather abruptly when the solar magnetic polarity flips. They attributed the change in slope to a systematic change in the interplanetary diffusion coefficient for cosmic rays.

In recent years, the McMurdo neutron monitor has been relocated to the South Korean Jang Bogo Station, and the mobile neutron monitor has been reconfigured. Therefore, the series of annual latitude surveys conducted between 1994 and 2007, as analyzed by \citet{NuntiyakulEA14}, cannot be extended directly. For this paper, we first adjusted the analysis using data from the NM at Mawson Station, 
 Antarctica instead of the McMurdo NM and added an analysis of two recent latitude surveys conducted in 2018 and 2019.




\section{Observations} \label{sec:data_col}
Data from several neutron monitors were used in this study. In this section, we briefly describe the monitors and the role they play in our analysis.

\subsection{Mobile Neutron Monitors}
\begin{figure}
    \centering
    \includegraphics[scale=1.2]{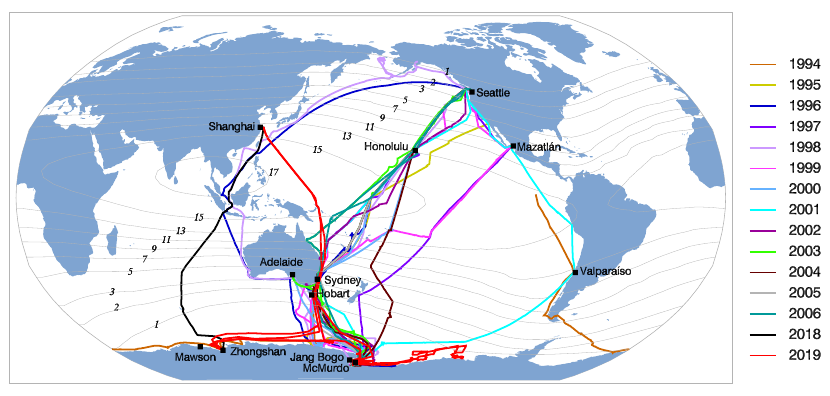}
    \caption{
    Tracks of ship-borne neutron monitor latitude surveys conducted during survey years 1994-2006 \citep{NuntiyakulEA14} and 2018-2019 \citep{KhamphakdeeEA21,Yakum2021} are shown with a contour map of the vertical cutoff rigidity (GV) calculated for February 11, 2019 at 12:00 UT.}
    \label{fig:route}
\end{figure}

All the latitude surveys actually began in one year and ended in another. We label all of them by the year in which they began and divide them into two groups, thirteen surveys from survey years 1994 to 2006 plus two in survey years 2018 and 2019. Figure \ref{fig:route} illustrates the routes taken by the monitors, distinguished by different colors.

To calculate the geomagnetic cutoffs we used the International Geomagnetic Reference Field (IGRF12) \citep{ThebaultEA2015}, for the Earth's main field, combined with a magnetosphere model by Tsyganenko \citep{Tsyganenko1987}. Cutoffs were calculated at the altitude of the Earth's tropospheric production ($\approx 20$ km).

The ``apparent cutoff rigidity'' utilized in this study considers both vertically and obliquely incident particles \citep{ClemEA97}. This is individually calculated at 1-hour intervals at the actual position of the ship using the time-dependent model of the magnetic field according to an efficient method \citep{BieberEA97}.

The apparent cutoff rigidity from the survey years 2018 and 2019 can be significantly higher than the vertical cutoff rigidity by 3.66\%. For instance, in the survey year 2018, the maximum apparent cutoff reached 18.08 GV, whereas the vertical cutoff was 17.43 GV. Similarly, in the survey year 2019, the maximum apparent cutoff was 16.71 GV, and the vertical cutoff was 16.11 GV.

\subsubsection{Thirteen Survey Years, 1994--2006}
Surveys from 1994 to 2007, documented by \citet{NuntiyakulEA14}, employed a standard three-detector neutron monitor (3NM64) housed within a shipping container referred to as the ``Tasvan.'' 
The term NM64 refers to the basic design \citep{Carimichael1964}, and ``3'' indicates using three neutron counters.
The Tasvan operated aboard the Polar Sea or the Polar Star (U.S. Coast Guard icebreakers) on approximately six-month voyages across the Pacific Ocean from Seattle in the USA to McMurdo, Antarctica, and back.

\subsubsection{Two Survey Years, 2018--2019}
For the survey years 2018 and 2019, the Tasvan was refurbished, reconfigured, and renamed the ``Changvan". 
We used a monitor similar to the 3NM64 used in the previous series but without the lead producer surrounding the central counter.  To be precise, the outer proportional counters, referred to as T1 and T3, are surrounded by lead rings, i.e., are leaded counters.
However, the central counter, called T2, is not surrounded by lead rings, i.e., it is an unleaded counter.
Thus we refer to the entire monitor as a semi-leaded neutron monitor, and it has an energy response distinct from an NM64. By operating these leaded and unleaded counters together, some spectral information can be obtained \citep{BieberEvenson91,BieberEA13}, while avoiding the systematic errors that can arise from comparing count rates from different locations, as depicted in Figure 2 of \citet{RuffoloEA16}. The accuracy of this method is dependent on understanding the relationship between the energy responses of different detector types. For the present study, we only use data from the two leaded counters.

\begin{figure}[!ht]
    \centering
     \begin{tabular}{@{}cc@{}}
    \includegraphics[width=0.51\textwidth]{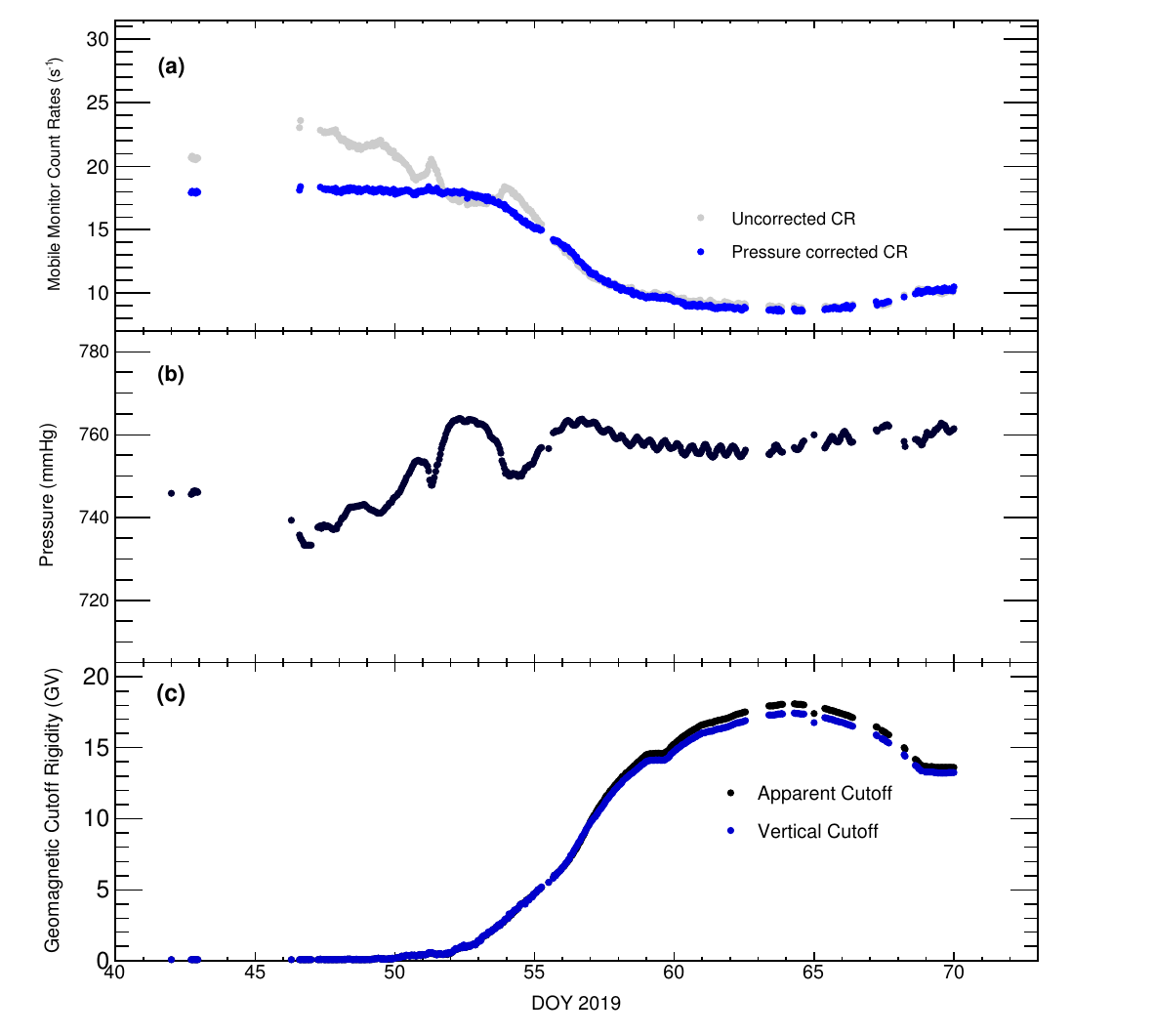}
    \includegraphics[width=0.51\textwidth]{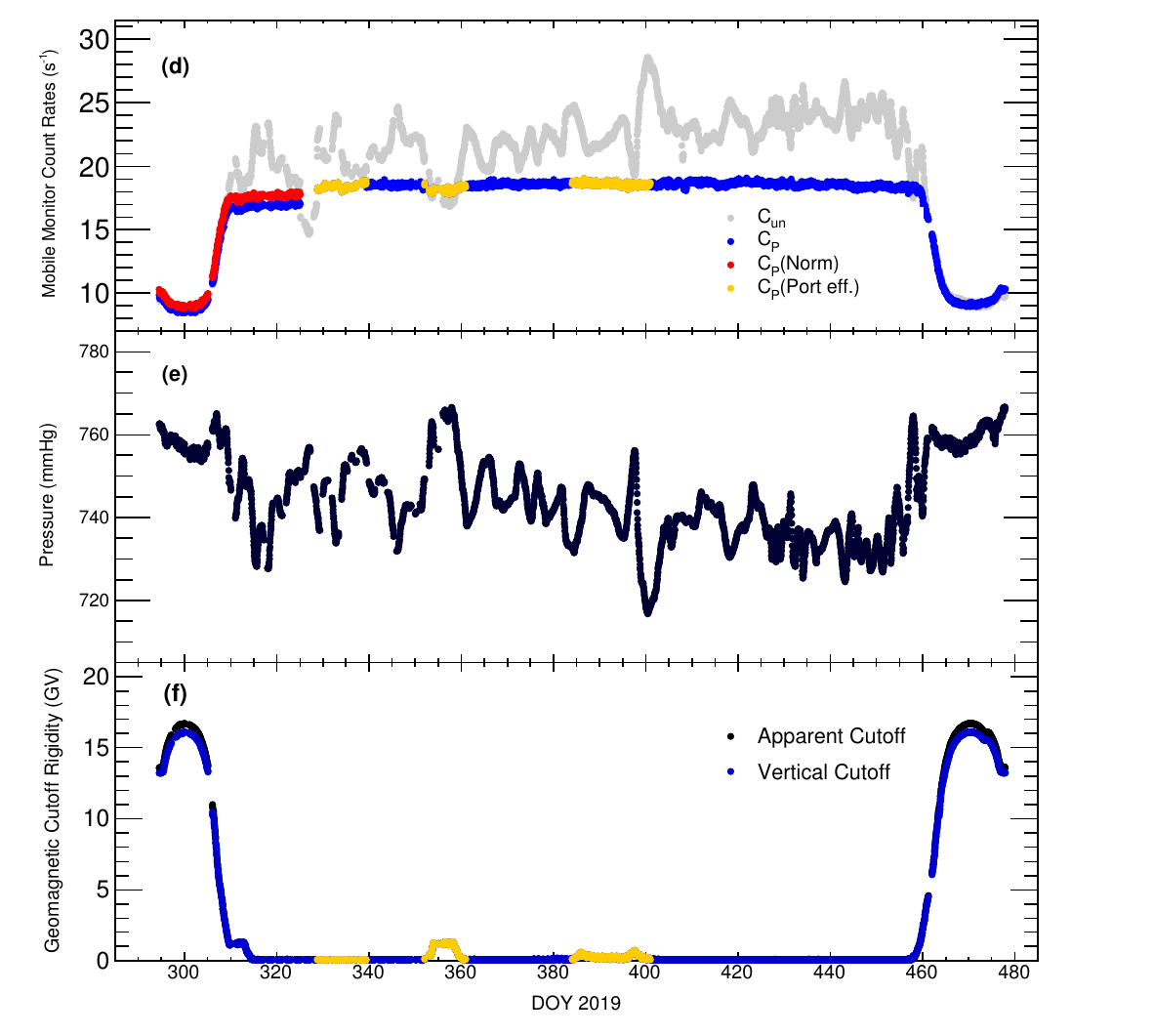}
    \end{tabular}
    \caption{ The hourly data for survey years 2018[(a)-(c)] and 2019 [(d)-(f)] as a function of time. Panels (a) and (d): Summed counting rates of the two leaded detector tubes. Gray points show the raw count rates, while blue points show pressure-corrected data. An attempt at normalizing the southbound data, shown in red, is discussed in the text. Panels (b) and (e): Barometric pressure. Panels (c) and (f):  Geomagnetic cutoff rigidity -- black for the apparent cutoff and blue for the vertical cutoff. Data shown in yellow were excluded from the analysis because they show evidence of environmental disturbance while the ship was in or near a port.}
    \label{fig:ChangV_CR}
\end{figure}

The Changvan operated onboard the icebreaker Xuelong between Shanghai and Zhongshan Station in Antarctica. During the 2018 survey year (2 November 2018 to 11 March 2019), the Changvan only operated on the return journey from Zhongshan Station to Shanghai from 11 February to 11 March 2019. Data were recorded during the 2019 survey year (21 October 2019 to 22 April 2020) both to and from Antarctica.  We used count rate data provided in \citet{KhamphakdeeEA21} for the complete survey data in 2018 and only the southbound data in 2019. We obtained additional data in the 2019 survey year from \citet{Yakum2021}. We present an overview of the data as a function of time in Figure \ref{fig:ChangV_CR} for both survey years.

\subsection{Fixed Neutron Monitor Stations}

\subsubsection{McMurdo and Jang Bogo}
McMurdo Station is a US Antarctic research station located at the southern tip of Ross Island (77.90$^\circ$S, 166.6$^\circ$E), with an altitude of 48 meters above sea level and a vertical cutoff rigidity of less than 0.3 GV. The Cosmic Ray Lab at the station housed an 18-tube NM64 neutron monitor comprising three sections with six neutron detector tubes each. The pressure-corrected count rate data used in this analysis were obtained from \citet{NuntiyakulEA14}, using a barometric coefficient ($\beta$) of 0.9866 $\%$/mmHg and a reference pressure of 730.00 mmHg. In 2015, the relocation of the McMurdo neutron monitor to the South Korean station at Jang Bogo, Antarctica, began with the removal of one section (six tubes) for shipment. Jang Bogo Station is 360 km from McMurdo Station.

From 14 November 2014, the McMurdo NM operated with only two sections (12 tubes) until 2017 when the final 12 McMurdo counters were shipped to Jang Bogo. The McMurdo monitor was completely shut down on 7 January 2017. By early 2018, all three sections were back together and operating at Jang Bogo. Initially, we had planned to use data from Jang Bogo to replace McMurdo to extend our analysis using the data from the recent two surveys. However, there was a data gap of about one month (from 5 January to 3 February 2020) and several months of frequent missing data after 19 February 2020. Due to these numerous data gaps, we instead used data from the Mawson NM. We did, however, refer to the Jang Bogo data (when available) to identify anomalies in the Mawson data.

\subsubsection{Mawson}\label{subsubMawson}
Mawson Station is located at the edge of the Eastern Antarctic plateau (67.60 $^\circ$S, 62.88 $^\circ$E) and is the oldest operational continental station south of the Antarctic Circle. The altitude is 30 meters above sea level, and the vertical cutoff rigidity is 0.22 GV.

From 1986 until October 16, 2002, the Mawson NM was a 6NM64. After October 17, 2002, the NM was upgraded to an 18NM64. In early 2020, the data acquisition system at Mawson was updated with new electronic firmware and computer software. Preparation of Mawson data is therefore complicated due to changes in the number of counter tubes in late 2002. To account for this change, we used a scaling factor of 2.851 to divide the 18NM64 data and normalize them to the 6NM64 count rate. We discuss details of this scaling in Section \ref{sec:ScalingMawson}.

\section{Data Reduction and Correction} \label{Sec:Data_Reduction}
We now discuss in some detail the reduction and correction of data for the latest two survey years, as well as data from the Mawson neutron monitor. For this discussion, all dates will be referred to as the ``day of year 2019'' (DOY 2019). We start counting on January 1, 2019 and continue thereafter. For the earlier surveys, we utilized pressure-corrected data directly from \citet{NuntiyakulEA14}.

\subsection{Changvan Data Cleaning} \label{Changvan:CleanningCR}
Interference of unknown origin obviously affected the Changvan counting rates, so we developed techniques to ``clean'' the data. Although we did not use the unleaded (T2) detector data in the final analysis, we did refer to these data in the cleaning process. The interference apparently occurred primarily in ``bursts'' so we were able to use the one-second resolution recording of the counts to advantage.

We, therefore, employed the technique discussed by \citet{NuntiyakulEA18} by generating histograms of 1-second readings for each tube for every hour. Since the Changvan count rates are much higher than the rates in that study, we used Gaussian rather than Poisson statistics to identify outliers of $\ge 30$ per second. We also removed ``frozen'' data where the exact same number of counts was recorded in each counter tube for at least three consecutive seconds. Finally, we removed seconds where all tubes recorded zero counts. After removing the bad seconds, the distribution appeared closer to a Gaussian distribution. We excluded any hour with less than 300 valid seconds of data as it was statistically unreliable. We then followed the steps outlined in Section 3 of \citet{NuntiyakulEA14} for each tube, either by deleting all tubes for that hour or excluding tubes that produced an unusual count rate. This data cleaning process was performed for both the semi-leaded monitor (Changvan) and Mawson NM.

Finally, we calculated the hourly average tube ratios (T1/T2, T2/T3, and T3/T1) for both surveys and plotted them on a single long plot as a function of time. Figure 4 in \citet{KhamphakdeeEA21} shows these ratios after correcting for the instrumental issues mentioned earlier. However, the ratios still had some outliers. For the count rate ratio T3/T1, we consider outliers to be beyond $\pm 3\sigma$ standard deviations from the mean value of the Gaussian distribution. For the count rate ratios T1/T2 and T2/T3, we consider outliers to be beyond $\pm 4\sigma$ from the mean value of the Gaussian distribution. We used a different threshold for ratios involving T2 because the unleaded counter is more sensitive to environmental changes. We only used data that fell within these limits for further analysis.

\subsection{Mawson NM Data Cleaning} \label{Mawson:CleanningCR}

Starting October 17, 2002, the number of counter tubes at Mawson increased from 6 to 18. The available data give the sum of count rates for the six tubes in each of the three sections, which are denoted as G0, G1, and G2. We cleaned the data using the ratios of total count rates for each section, comparing the section ratios with the average section ratios for the entire survey year. The averaged count rates for sections 0, 1, and 2 were denoted as A0, A1, and A2, respectively. In case the count rate of a specific section was unavailable, we calculated the corrected count rate by [(A0+A1+A2)/(A1+A2)]x(G1+G2). If two sections were removed, the remaining section was used to calculate the corrected data. If none of the tubes were operational, the count rate was recorded as a data gap.

\subsection{Atmospheric Pressure Correction}\label{subsec:pressurecor}
Figure \ref{fig:ChangV_CR} illustrates the strong negative correlation between a NM count rate  and barometric pressure. To accurately estimate the response functions for cosmic rays, it is necessary to make significant corrections for barometric pressure variations. One feature of the NM is that nearly all of the variation in response is determined by the simple air mass measured by barometric pressure. For stationary installations, a simple approach is to plot changes in the natural logarithm ($\ln$) of the count rate against local barometric pressure over a long period. Since these processes are fundamentally uncorrelated, a sufficiently long time series results in an accurate correction that can be applied to the installation. At Mawson, we applied a barometric coefficient of $\beta$~=~0.9439 \%/mmHg with a reference pressure of 742.56 mmHg for pressure correction of the counting rate.

The correction for the Changvan is more difficult since there is a strong correlation between atmospheric properties and geomagnetic cutoff. In the Arctic and Antarctic regions, for example, the average atmospheric pressure is around 20 mmHg lower than at the equator. The atmospheric structure is also different. Various techniques have been developed to organize and interpolate neutron monitor data and calculations to address these issues. Using the 13 Tasvan surveys \citep{NuntiyakulEA14}, were able to determine a rigidity dependent $\beta = (1.006\times10^{-2} - 1.53\times10^{-4}P_c)$ \%/mmHg, where $P_c$ is the apparent cutoff
rigidity in units of GV. The most recent surveys are consistent with this but add little to the statistical accuracy. Therefore we used this $\beta$ with the exponential function $C_p = C\exp[\beta(p-p_\mathrm{ref})]$, where $C_p$ is the pressure-corrected counting rate, $C$ is the uncorrected counting rate, $p$ represents the barometric pressure in units of mmHg, and $p_\mathrm{ref}$ denotes the reference pressure. For the Changvan monitor, we used $p_\mathrm{ref}=760$ mmHg.

\subsection{Changvan Data Removal} \label{subsec:dataremovalandnorm}

During the two-year period of Xue Long surveys, we encountered occasional gaps in the data collection due to technical difficulties. Specifically, we observed a significant decrease in count rates during DOY 325--328 (November 21--24, 2019) during the second survey year (as shown in Figure \ref{fig:ChangV_CR}(d)) when the ship was docked at Zhongshan station to load cargo. We attributed this decrease to the instrument being covered by other cargo containers during the unloading. To avoid using contaminated data, we removed the affected data from further consideration.

Even after this removal, there is a clear discontinuity in the count rate even though there is minimal change in solar modulation during the gap (as seen in Figure \ref{fig:solar_modulation}). We conclude that the shift occurred due to changes in the ship's configuration. During the 2019 survey year, the ship was heavily loaded during its journey to Antarctica but nearly empty during its return to Shanghai. The Changvan was carried with the other containers, not separately, on an open deck.

To account for this, we searched for a simple, multiplicative normalization factor for the count rate. The factor that equalizes the rates  at Zhongshan causes a significant mismatch at a higher cutoff. Looking in detail at the course plots, we found that the southbound and northbound paths overlapped exactly as the ship crossed the Equator at an apparent cutoff rigidity from 16.30 GV to 16.35 GV. We used this period to calculate count rate ratios between the southbound and northbound legs and obtain normalization factors for the three counter tubes. For T1, T2, and T3, the normalization factors were found to be 1.06022, 1.04353, and 1.03016, respectively. The red points in Figure \ref{fig:ChangV_CR}(d) show the result of this attempt. The discontinuity is somewhat reduced but not eliminated.

The implication of this normalization failure is that the yield function of the Changvan was significantly altered by the presence of the other containers on the southbound segment. We, therefore, have not used the southbound segment in further analysis. In our future work with the semi-leaded detector in the Changvan, we also intend to investigate the altered yield function more thoroughly as it may provide yet another means of investigating spectral changes.

\section{Data Normalization} \label{Sec:Scaling}

This study uses data from several NMs with changing environments. To compare the different datasets, we must adjust the measured count rates to simulate identical monitors in a constant environment. To do this, we first correct the count rates for atmospheric pressure variations on an individual basis, as discussed above. Then we make the assumption that the yield functions of all the monitors are identical except for a scaling factor. As noted above, this assumption can fail, but historically it has proved acceptable for standard NM64 monitor units at nearly the same altitude. The best example of this is the tendency of the end detectors in a neutron monitor to have significantly lower counting rates than the central detectors without any noticeable change in the yield function other than the overall amplitude. Our adjustments, therefore, consist of determining these scaling factors using data when the detectors are simultaneously at similar geomagnetic cutoffs.

\subsection{Mawson Configuration Change}\label{sec:ScalingMawson}

\begin{figure}
	\centering
	\includegraphics[scale=0.9]{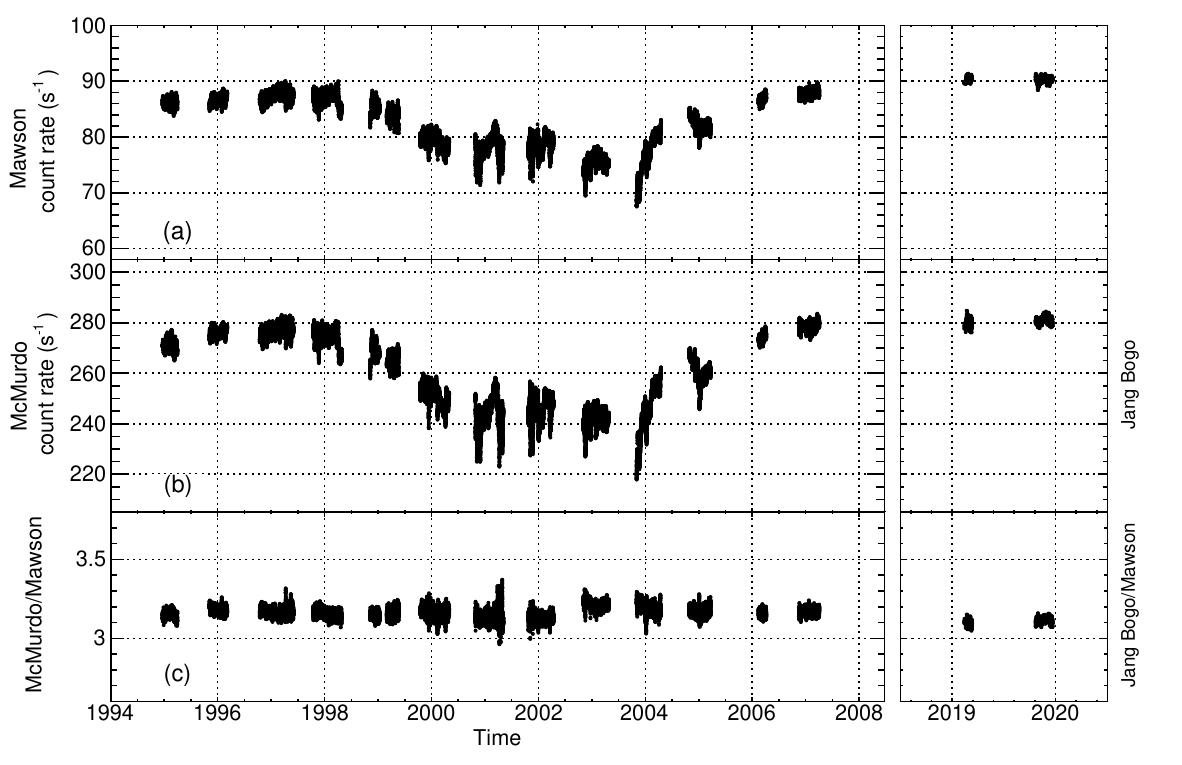}
	\caption{Panel (a): Pressure-corrected hourly count rate of the Mawson neutron monitor for all survey years, after adjustment for the configuration change in 2002. Panel (b): McMurdo and Jang Bogo hourly count rates for the same period. Panel (c) Ratio of the data in (b) to data in (a).}
	\label{fig:McMurVsMawson_ratio}
\end{figure}

Figure \ref{fig:McMurVsMawson_ratio} compares the counting rates of the monitors at McMurdo, Mawson, and Jang Bogo at the times of all the latitude surveys in our study. For comparison with the mobile monitors, we use the Mawson data in all cases. To ensure the accuracy of our data, we used the count rate ratio between Mawson Station and McMurdo Station to remove any anomalies. We identified outliers using the interquartile range (IQR) method, where we found the median (middle value) of the lower and upper halves of the data, known as quartile 1 (Q1) and quartile 3 (Q3), respectively. The IQR is the difference between these two quartiles, and we removed any Mawson count rates that fall below Q1 $-$ 1.5IQR or above Q3 $+$ 1.5IQR.

After removing the outliers, we split the data into two periods, before or after the number of counter tubes was increased from 6 to 18 in 2002. After 2002 we applied a scaling factor of 2.851, which we derived from the count rate ratio before and after the upgrade.

The McMurdo NM was permanently closed in 2017, and its neutron counters were gradually transferred to Jang Bogo station. Therefore, for the recent two survey years (2018-2020), we used pressure-corrected Jang Bogo count rates for outlier removal.

\subsection{Tasvan and Changvan}\label{sec:two_latSur}

We also assume that the Tasvan and Changvan are similar enough that the counting rate of the Changvan, $R_C$, would be related to the rate of the Tasvan, $R_T$, by a universal scale factor $S$. In an ideal situation where the configurations and environments were identical, $S$ would be an absolute constant that could be determined once and for all by running the two ``side-by-side'' for a time interval long enough to get good statistics and then taking the ratio of the count rates. Unfortunately, the environments change year to year, and, as the Changvan is the reconfigured Tasvan, even in principle they never could operate together.

\begin{figure}
	\centering
	\includegraphics[scale=0.9]{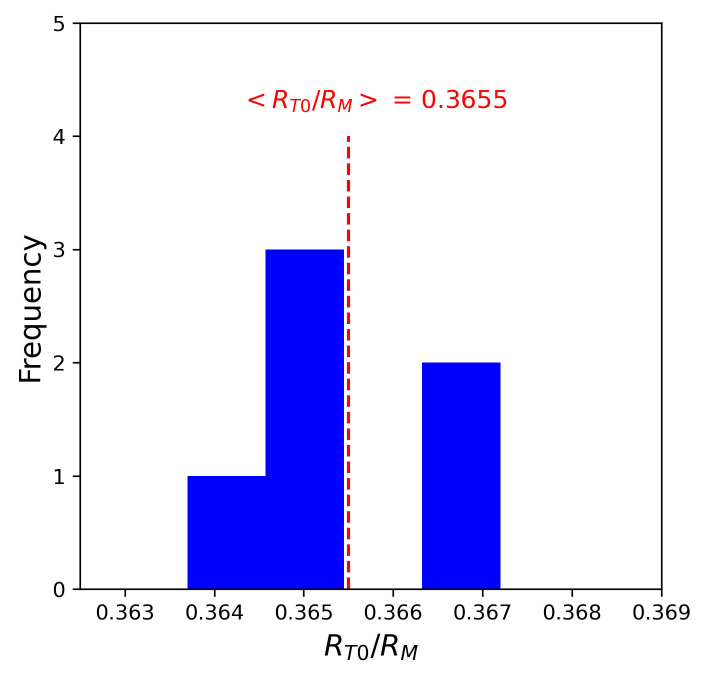}
	\caption{Distribution of the Tasvan (at zero cutoffs) to Mawson count rate ratio for annual latitude surveys in positive solar magnetic polarity}
	\label{fig:TasMawson}
	
\end{figure}

We also assume that the counting rate of either $R_{T0}$ or $R_{C0}$ at zero cutoff is proportional to the counting rate of the Mawson NM, $R_M$. In other words, $R_{T0} = Y_T R_M$ and $R_{C0} = Y_C R_M$. As $S$ is assumed to be universal, it can be computed at zero rigidity. With these definitions, $S = Y_T/Y_C$ and the $Y$ are determined by $Y_T = R_{T0}/R_M$ and $Y_C = R_{C0}/R_M$.

At this point, we must proceed differently for the Tasvan and Changvan. For the Tasvan, we have multiple surveys, with counting rates already normalized for configuration changes by \citet{NuntiyakulEA14}. When expressed as ratios to Mawson, the data for positive solar magnetic polarity are displayed in Figure \ref{fig:TasMawson}. Since these measurements were already corrected for configuration changes, the variability in the figure results from the difference in averages for McMurdo and Mawson. With different asymptotic directions, it is not surprising that the average counting rates are not exactly proportional to each other. We adopt the average of these values, 0.3655, as $Y_T$. With only two Changvan surveys, it is not possible to separate survey-to-survey variation from an average value. Formally one could define an average of the two numbers and calculate the deviation of each from the average, but that is the same as just computing the values of the ratios separately. Hence we compute $Y_S$ for survey years 2018 and 2019 to be 0.2004 and 0.2035, respectively. This leads to corresponding determinations of $S$: 1.824 and 1.795.

\section{Linear Correlation of Neutron Monitor Counting Rates}\label{sub:LinCor}

\begin{figure}
    \centering
    \includegraphics[width=0.95\textwidth]{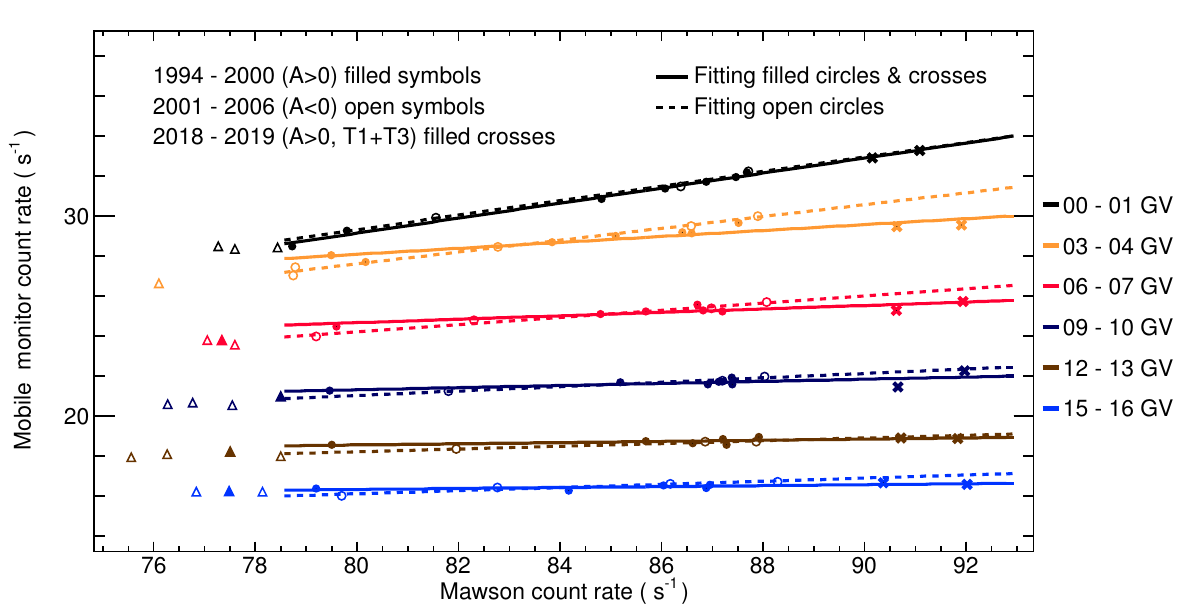}
    \caption{Regression of the mobile neutron monitor count rate in different apparent cutoff rigidity bins against the count rate of the Mawson NM. To improve clarity, we plotted every third bin. Filled circles indicate positive (A$>$0) solar magnetic polarity, while open circles indicate negative polarity.
    Lines are fit separately for positive and negative polarity. Triangles indicate times of uncertain polarity not included in the fits.}
    \label{fig:MWvsMobile13y}
\end{figure}

An important finding of \citet{NuntiyakulEA14} was that the variation with solar modulation of any two NM counting rates exhibits a linear correlation, with a regression slope dependent only on solar magnetic polarity. This is shown in Figure~\ref{fig:MWvsMobile13y}, which is based on Figure 10 of \citet{NuntiyakulEA14}. The main difference is that the horizontal axis is the Mawson counting rate as opposed to McMurdo.

Each point is produced as an average, over one annual latitude survey, of the counting rate of the mobile monitor whenever it was at a particular cutoff rigidity ``bin'' of width 1GV, plotted at the average counting rate of Mawson -- the average in each case being taken to correspond in time to the mobile observations. Since the level of modulation varied over each survey, the points for a given survey do not line up vertically. Points for the new surveys have been added by scaling the Changvan data by the factors discussed above. For clarity, only every third rigidity bin is plotted in the figure. Data for positive solar magnetic polarity (A$>$0) are indicated by filled symbols and data for A$<$0 by open symbols. We excluded data for times of uncertain polarity, indicated by triangles, which were data of the survey year 2000, and all data with the  Mawson count rate below 78.5 s$^{-1}$.

The regression for each bin against the Mawson count rate confirms the change in slope before and after the solar polarity reversal in the year 2000, as reported in \citet{NuntiyakulEA14}. Our determination of the scaling factors is essentially equivalent to forcing the zero rigidity points to lie on the line with the others, but with this normalization, the new data at other rigidities fall nicely in line with the previous observations during the same magnetic polarity. This is true even though the modulation level is significantly less (Mawson rates are higher) during the new surveys.

\section{Response Function}\label{sub:resp}

The spectral crossover itself is best seen by comparing differential response functions at solar minimum. The differential response function is the derivative of the integral response function, which in turn is simply the counting rate of the mobile monitor as a function of cutoff rigidity. Ideally, the entire survey should be conducted with a constant modulation level, but in reality, the modulation level is not constant over the survey. For Figure~\ref{fig:MWvsMobile13y} this does not matter, but for the response function, it is necessary to apply a correction for this varying ``short term'' modulation.

\begin{figure}
	\centering
	\includegraphics[scale=0.55]{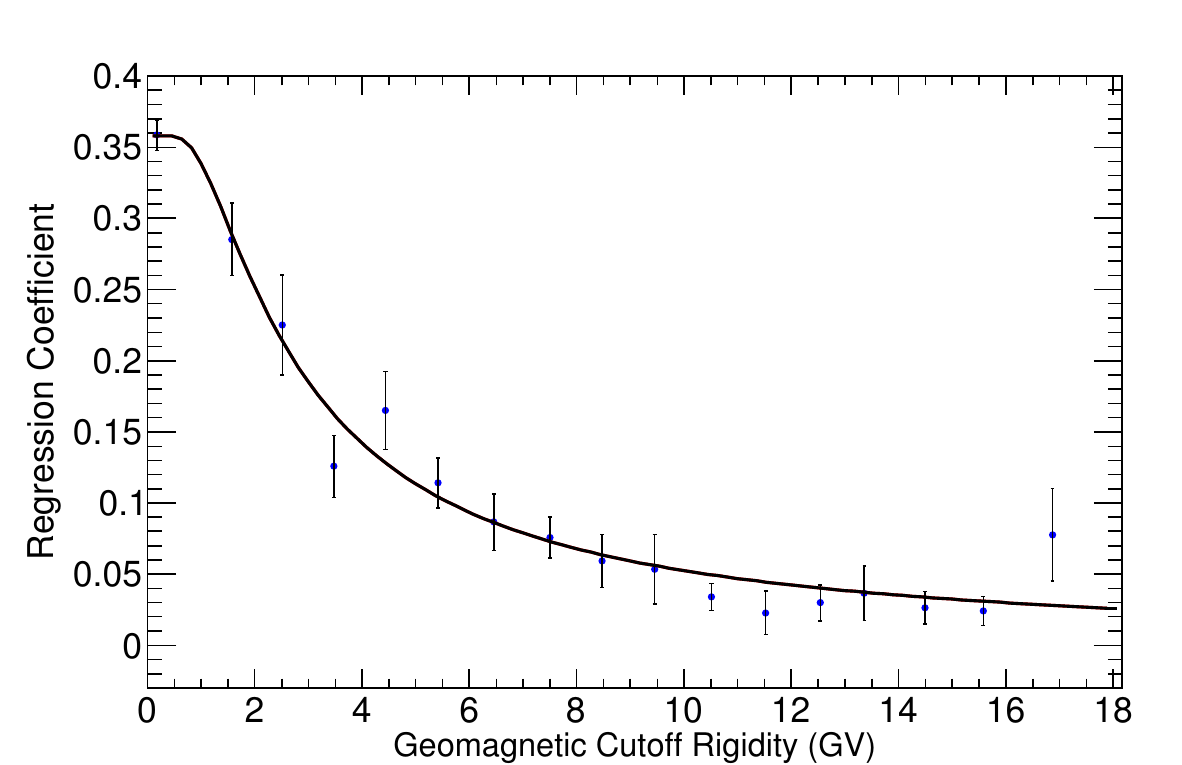}
	\caption{Regression coefficients (slopes) of mobile monitor vs. Mawson NM count rate plotted as a function of geomagnetic cutoff rigidity (GV).}
	\label{fig:reg_coef}
\end{figure}

\subsection{Short Term Modulation}\label{sec:ShortTerm}

To correct for short-term modulation, we use another result from \citet{NuntiyakulEA14}, namely, both long and short-term modulation follows the same regression line. An exception to this is during Forbush decreases, so data for these time intervals have been removed in cleaning the data from the various surveys.

In Figure~\ref{fig:reg_coef}, we show the regression coefficients (slopes) for the lines described in Figure~\ref{fig:MWvsMobile13y}. The highest rigidity bin is actually 3 GV wide for better statistics. We plot the coefficient for each rigidity bin, not every third bin, as in Figure~\ref{fig:MWvsMobile13y}. The points are plotted at the average (not central) rigidity. Error bars are assigned based on the scatter of the points from the fit line. Although the rigidity dependence is quite systematic, there are significant statistical fluctuations. For the survey, we want to correct each hour of data individually and bin the data according to the cutoff calculated for that hour. We, therefore, use the fit line shown in Figure~\ref{fig:reg_coef} to smooth and interpolate the regression coefficient. The fit is to a function given by Eq.~(\ref{Eq:regression(S)}) with three parameters ($S_0$, $\epsilon$, and $\zeta$).

\begin{equation}
    S_m = S_0\left(1 - e^{-\epsilon P_c^{-\zeta} }\right),
    \label{Eq:regression(S)}
\end{equation}

This function is in the Dorman form (see below), which has the property that it provides a plateau region at a low cutoff with a rather sharp transition to the fit at a higher cutoff. Below approximately 2 GV, the cutoff is atmospheric, not geomagnetic, and no variation with geomagnetic cutoff is expected. The fit yields parameter values $S_0 = 0.358$, $\epsilon = 2.915$, and $\zeta = 1.264$.

Using the value of $S_m$ as a function of $P_c$, we were able to correct for short-term modulation variations by applying Eq.~\ref{Eq:short-termCorr}:

\begin{equation}
R_{mob} = R_{CP} - S_m(R_M - \overline{R_M}).
\label{Eq:short-termCorr}
\end{equation}

Here, $R_{mob}$ is the count rate of the mobile monitor, which has been corrected for both pressure and short-term variations. $R_{CP}$ is the normalized Changvan rate corrected for pressure, $R_M$ is the counting rate at Mawson at the time of the observation, and $\overline{R_M}$ is the Mawson rate averaged over the survey.

\subsection{Integral and Differential Response Functions}\label{sec:SIntDiff}

\begin{figure}
    \centering
    \includegraphics[scale=0.7]{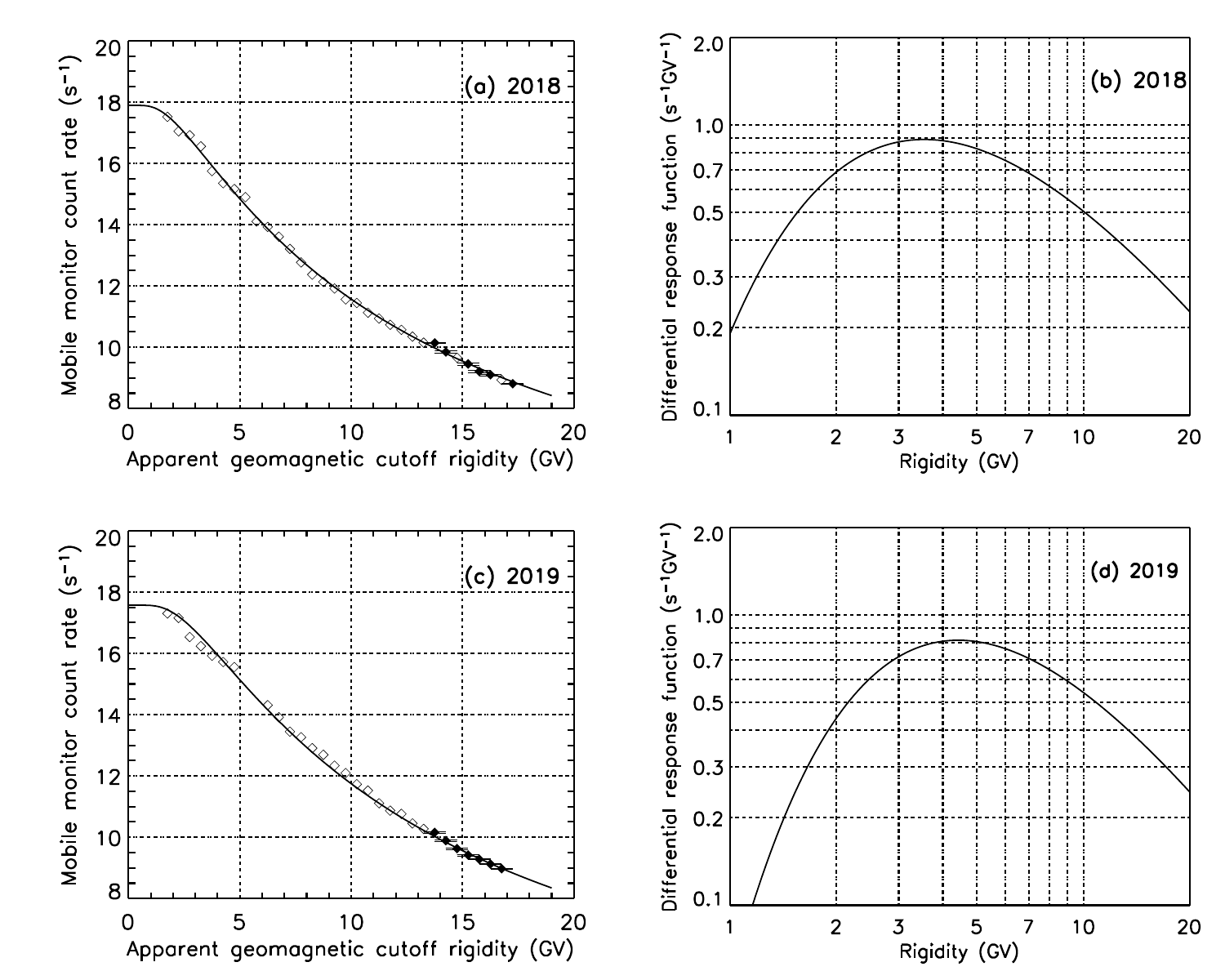}
    \caption{Response functions for survey years 2018 and 2019. Panels (a) and (c) show the data and the Dorman function fits, determining the integral response functions. Panels (b) and (d) are the differential response functions calculated analytically from the Dorman functions. A filled symbol indicates the availability of data for multiple voyage segments. The vertical error bar signifies the standard error across these segments; in all cases, the error bar is smaller than the plotted symbol. An open symbol with no error bar indicates that data were available for only one segment.}
    \label{fig:response}
\end{figure}

The relationship between the counting rate ($N$) and the geomagnetic cutoff rigidity ($P_c$) shown in Figure~\ref{fig:response}, panels (a) and (c), for the 2018 and 2019 surveys is the integral response function. Data are plotted as averages over 0.5 GV rigidity bins. We fit the data to the Dorman  function \citep{Dorman70} given by Eq. (\ref{Eq:DormanN}) where $N_0$, $\alpha$, and $\kappa$ are free parameters: 

\begin{eqnarray}
N &=& N_0(1 - e^{-\alpha P^{-\kappa}_c}), \label{Eq:DormanN} \\
DRF &=& N_0 \alpha P^{-\kappa-1} \kappa (e^{-\alpha P^{-\kappa}}).
\label{Eq:Dorman}
\end{eqnarray}
\noindent The fit spans 0.15 GV to 18 GV, corresponding to the range used by \citet{NuntiyakulEA14}.
We used the Levenberg-Marquardt algorithm to minimize the chi-squared values. Table \ref{tab:DormanParameters} shows the three parameters for each survey year.
Except for the normalization $N_0$, the parameters have no physical interpretation, but the fitting allows an analytic determination of the differential response functions given by Eq. (\ref{Eq:Dorman}), as shown in Figure~\ref{fig:response}, panels (b) and (d).

\begin{table}\label{tab:DormanParameters}
\centering
\caption{The Dorman Parameters}
\label{Tab:Dorman_Parameters}
\begin{tabular}{|c|c|c|c|}
\hline
Survey year  &$N_0$& $\alpha$ & $\kappa$ \\ \hline
2018         & 17.89 & 6.07   & 0.766    \\ \hline
2019         & 17.57 & 7.64   & 0.840    \\ \hline

\end{tabular}
\end{table}

\section{Crossover Results and Conclusions}

\begin{figure}
    \centering
    \includegraphics[width=\textwidth]{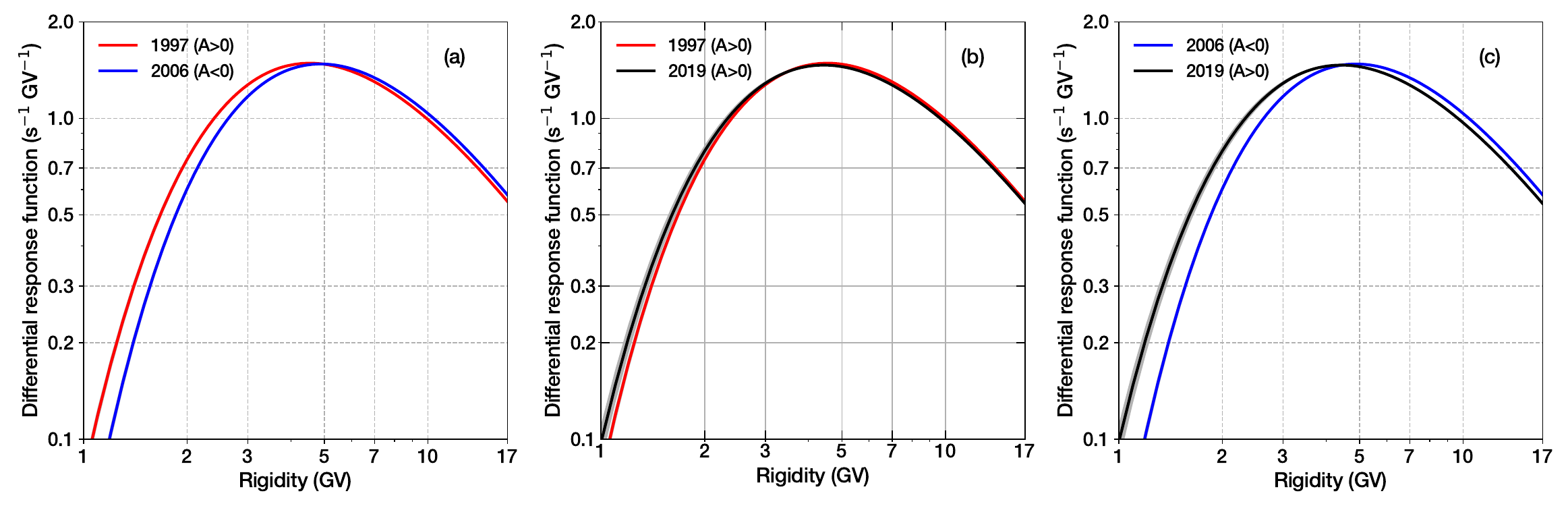}
    \caption{Differential response functions for survey year 2019 compared to earlier surveys performed near solar minima. Panel (a): The 1997 (red) and 2006 (blue) surveys, which had opposite solar magnetic polarity, show a crossover near 5 GV. Panel(b): The 1997 (red) and 2019 (black) surveys for T1 and T3, which had the same polarity, do not show a crossover. Panel (c): Results of the 2006 (blue) and 2019 (black) surveys, again with opposite polarity, show a crossover near 5 GV, similar to (a). The solid line represents the best-fit line, and the shaded area represents the possible range of fits ($\pm 2 \sigma$).}
    \label{fig:crossover}
\end{figure}

Figure~\ref{fig:crossover} displays the differential response function for the 2019 survey year, along with previous results reported in \citet{NuntiyakulEA14}. To represent the scatter in the data, we also included statistical errors in error bands in Figure \ref{fig:crossover}, although the bands appear as broad lines on the scale of the figure. We only used data from the 2019 survey, not 2018. This choice is based on stronger indicators in Figure \ref{fig:response}, showing improved segment representation. Data gaps are significantly reduced in 2019, particularly during high geomagnetic cutoff periods, as seen in figures \ref{fig:ChangV_CR}(a) and \ref{fig:ChangV_CR}(d). The new results continue the historical pattern of spectral crossovers.  The crossover is clearly visible when the solar magnetic polarities are opposite. In contrast, there is no evidence of the crossover when observing the same positive magnetic polarity.

The overall modulation level in the 2019 survey is less than that in the 1997 survey. This is seen mostly at lower rigidity, where the $DRF$ for 2019 lies noticeably above that of 1997. The lower level of modulation (i.e., higher GCR flux) is consistent with the lower sunspot number in 2019, as can be seen from Figure 1.

In addition to confirming the crossover for the most recent solar minimum period, we showed that the linear analysis of the mobile monitor counting rates against the fixed Mawson NM continues to have a slope that changes with solar magnetic polarity. The additional surveys support the proposal by \citet{NuntiyakulEA14} that different magnetic polarity effects are operating in solar modulation -- drifts dominating at lower rigidity and helicity-modulated diffusion at higher rigidity. This is also consistent with the results of \citet{MangeardEA18}, who presented the first observations of solar modulation over a sunspot cycle from a neutron monitor at high cutoff rigidity ($\sim$17 GV).  
That work found that at high rigidity, the GCR flux variations are anticorrelated with the interplanetary magnetic field and thus should be dominated by diffusion, in contrast with solar modulation at lower cutoff rigidity, where other factors such as drifts play an important role.

\section{Acknowledgements}

The 2018 and 2019 surveys were conducted in cooperation with the Polar Research Institute of China (PRIC). We gratefully acknowledge the logistical support provided by Australia's Antarctic Program for operating the Mawson neutron monitor. This research was supported by the Development and Promotion of Science and Technology Talents Project (DPST), NSRF via the Program Management Unit for Human Resources \& Institutional Development, Research, and Innovation [grant number B39G660028], and the National Science and Technology Development Agency (NSTDA) and National Research Council of Thailand (NRCT): High-Potential Research Team Grant Program (N42A650868). We also extend our appreciation to the Northern Science Park (Chiang Mai) for providing laboratory space that enabled our research team to function smoothly. Additionally, we are thankful to the ITSC of Chiang Mai University for providing us with an on-demand server to facilitate the processing of cutoff rigidity calculations.

%

\vspace{5mm}










\end{document}